\pdfoutput=1
\documentclass[aps,10pt, prl,superscriptaddress,floatfix,twocolumn,showpacs]{revtex4-2}
\usepackage[colorlinks=true,linkcolor=blue,citecolor=blue]{hyperref}
\usepackage{amsfonts}
\usepackage{amsmath}
\usepackage{mathrsfs}
\usepackage{amssymb}
\usepackage{amsbsy}
\usepackage{epsfig}
\usepackage{graphicx}
\usepackage{epstopdf}
\usepackage{mathdots}
\usepackage{color}
\usepackage{cleveref}
\usepackage{physics}
\usepackage{bm}
\makeatletter

\newcommand{\Rmnum}[1]{\expandafter\@slowromancap\romannumeral #1@}
\makeatother

\begin{document} 
\title{Non-Hermitian second-order topological superconductors}

\author{Xiang Ji}
\affiliation{Department of Physics, Jiangsu University, Zhenjiang, 212013, China}

\author{Wenchen Ding}
\affiliation{Department of Physics, Jiangsu University, Zhenjiang, 212013, China}

\author{Yuanping Chen} \altaffiliation{chenyp@ujs.edu.cn}
\affiliation{Department of Physics, Jiangsu University, Zhenjiang, 212013, China}

\author{Xiaosen Yang} \altaffiliation{yangxs@ujs.edu.cn}
\affiliation{Department of Physics, Jiangsu University, Zhenjiang, 212013, China}
\date{\today}

\begin{abstract}
The topology of non-Hermitian systems is fundamentally changed by the non-Hermitian skin effect, which leads to the generalized bulk-boundary correspondence. Based on the non-Bloch band theory, we get insight into the interplay between the non-Hermiticity and the second-order topological superconductors in two spatial dimensions. We investigate that the non-Hermiticity drives both the bulk states and topological edge modes to accumulate toward opposite corners of the system depending on the particle and hole degrees of freedom protected by the particle-hole symmetry. Furthermore, the degeneracy of the Majorana corner modes can be broken in terms of both the eigenenergies and the eigenstates. Through an edge theory analysis, we elucidate the impact of non-Hermiticity and enable the extension of higher-order topological superconductors to the realm of non-Hermitian systems.  We show that $Z_2$ skin effect and $Z_2$ skin-topological modes reveal the universal characteristics of non-Hermitian second-order topological superconductors and the generalized bulk-boundary correspondence is further enriched by the particle-hole symmetry. 
\end{abstract}

\maketitle

\section{Introduction}
Topological superconductors (TSCs) have attracted great interest in the past two decades for their potential applications in fault-tolerant topological quantum computation~\cite{prb2000readPaired,kitaev2001unpaired,prl2001gorkovSuperconducting, IvanovPhysRevLett2001, nayak2008non, prb2008schnyderClassification, prl2009qiTime, hasan2010colloquium, Sternnature2010, qi2011topological, Bernevig+2013, deng2016majorana}. 
The hallmark characteristic of the TSCs is their bulk-boundary correspondence: the emergence of robust Majorana zero modes (MZMs) is the consequence of the topologically nontrivial bulk~\cite{fu2008PhysRevLett, Leijnse_2012, jpsj2012tanakaSymmetry, sato2017topological, Science2018PZ_Observation, Science2018DW_Evidence, prb2019tamnuraOdd}. The bulk-boundary correspondence is enriched by a new class of TSCs which is dubbed as higher-order topological superconductors~\cite{prb2018tunable, prb2018majorana,prl2018wangHT-MCS, PRB_2019_ZBY, prb2019ghorashi,prl2019volpez, prl2020wu, prr2020zhangTopological, prb2021ghosh, prb2023chewHOTS}. In contrast to conventional first-order TSCs, an $n$th-order TSC in $d$ spatial dimension hosts $(d-n)$-dimensional MZMs ($1<n\leq d$)~\cite{PRL2019ZBY_odd-parity,xie2021higher,PhysRevB.97.205136, PRR2020Tiwari_Chiral}. 
A surge of realizations of the higher-order topological superconductors have been proposed on various experimental platforms~\cite{PRL2018ZBY_Majorana,PRB2018wangWeek,PRL2019XYZhu, PRL2019RXZhang_Helical,prl2019zhangHOtopology,PRB2020Bitan_dirac,PRB2021Luo_weak}.

Very recently, the topology has been extended into the non-Hermitian  systems~\cite{benderPhysRevLett1998,GongPhysRevX2018,EzawaPhysRevB2019,lee2020unraveling,KawabataPRX2019,EdvardssonPhysRevB2019,avila2019non, prl2020yangSpin,APS2020XueDirac,zengPRB2020,ashida2020non,Kocharxiv2020,wuhongPhysRevB2020,BergholtzRevModPhys2021,prl2021YangExceptional,ding2022non,lin2023topological}. The non-Hermitian topological systems have aroused lots of interest for the novel topological properties without Hermitian counterparts~\cite{TonyPhysRevLett2016,ZhouPRB2018,LonghiPRL2019,YangPRB2019,HochenPhysRevResearch2020,kouPhysRevB2020,kouPhysRevB2020b,BorgniaPRL2020,zhesenPhysRevLett2020, HelbigNaturePhysics2020, xiao2020non, Kawabata_2020, ZhangNC2021a, wang2022amoeba}. In particular, an intriguing phenomenon is reported, known as the non-Hermitian skin effect ~\cite{PRL2018Wang_Edge,PRB2018Martinez,PRL2019Wang_Real,PhysRevLettwangzhong2019a,zhangPRL2020,li2020critical,zhang2022universal}. The exponential localization of the eigenstates profoundly enriches the bulk-boundary correspondence~\cite{PRL2018Wang_Chern,MartinezAlvarez2018,ShenPRL2018,YokomizoPhysRevLett2019,LeePRB2019,prl2019LeeAnomalous,prb2021yangNHBBC} and expands the horizon for the study of topological phases of matter, therefore igniting extensive interest~\cite{PRL2019TL_second,prl2020Budich,PRA2020LiuCircuit,Guo2021PhysRevLett,prl2021wanjura_correspondence,prl2021HughesGeometric,PRL2022Longhi_self-healing,prl2022wangNHEB,prl2022yanDynamic,PhysRevBNHHOSC2022}. In addition, the non-Hermitian skin effect can induce skin-topological effect in the topological edge modes~\cite{LeePhysRevLett2019, LinhuPhysRevLett2020, prb2020higherorder, zou2021observation, prb2021fuNHSOSKM, prb2022gongHybrid,ma2024PRRNon}. Thus, the non-Hermitian skin effect and the topological properties of the non-Hermitian systems can be immensely enriched by symmetry and dimension. 
However, the interplay between the non-Hermiticity and higher-order topological superconductivity with particle-hole symmetry (PHS) remains unclear and needs to be uncovered.

In this work, we generalize the second-order TSCs into two-dimensional (2D) non-Hermitian systems to investigate the novel phenomena induced by the non-Hermiticity of the second-order topological superconductors with PHS. In contrast to the Hermitian cases, the bulk eigenstates and the topological edge modes of the non-Hermitian second-order topological superconductors (NHSOTSCs) exhibit non-Hermitian skin effect and localize at the corner of the system. By employing the non-Bloch band theory~\cite{wang2022amoeba} and the current function~\cite{zhang2022universal}, we deduce that the non-Hermitian skin effect corresponds to the $Z_2$ skin effect protected by PHS~\cite{PRL2019Wang_Real, JPCM2023Yang_Universal}, with corresponding eigenstates of particle and hole localizing at opposite boundaries. The $Z_2$ skin effect induces the topological edge modes to accumulate toward the opposite corners, which can be dubbed  ``$Z_2$ skin-topological modes". Furthermore, the eightfold degeneracy of the Majorana corner modes (MCMs)  is broken into four MCMs with pure imaginary energies and four MZMs for a $d$-wave NHSOTSC with time-reversal symmetry (TRS). The MCMs with pure imaginary energies and the MZMs are localized at different corners. Subsequently, we study the topological phase transitions between the NHSOTSC and the topologically trivial superconductor. Utilizing the non-Bloch band theory, we determine that the energy gap closure corresponds to Bloch points~ \cite{PRL2019Wang_Real, JPCM2023Yang_Universal}. We employ edge theory~\cite{PRL2018ZBY_Majorana,PRL2019TL_second} around Bloch points to gain an intuitive understanding of the NHSOTSC phase, offering a methodology to extend the higher-order topological superconductors from Hermitian to non-Hermitian systems.

\section{Majorana Corner Modes and $Z_2$ skin effect}

\begin{figure}
\centering     
\includegraphics[width=0.45\textwidth]{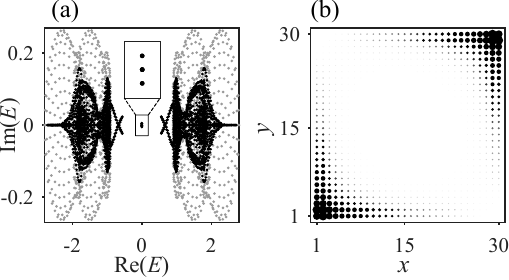}
\caption{(a) The energy spectrum for a $30 \times 30$ square lattice under OBC (black dots) and PBC (gray dots), respectively. 
The energies around zero are shown in the inset.
(b) The summed probability distributions of the bulk states which are localized at opposite corners of the lattice.
Parameters are set as $t_x=t_y=1$, $A_x=A_y=1$, $\mu=0.4$, $m_0=0.4$, $\delta_x=\delta_y=0.2$ and $\Delta_x=-\Delta_y=0.5$.}
\label{fig_bulk}
\end{figure}

We start from a 2D non-Hermitian superconductor model whose Bloch Hamiltonian can be written in the Bogoliubov-de Gennes (BdG) form:
\begin{equation}
H(\mathbf k)=
\mqty(H_{N}(\mathbf k)-\mu &\Delta(\mathbf k) \\  \Delta^T(-\mathbf k)&\mu -H_{N}^T(-\mathbf k)),
\label{eq_1}
\end{equation}
where $H_N(\mathbf k)=M(\mathbf k)\sigma_z + A_x \sin k_x s_z\sigma_x + A_y\sin k_y\sigma_y$ with $M(\mathbf k) = m_0 - t_x\cos k_x - i\delta_x\sin k_x - t_y\cos k_y - i\delta_y\sin k_y$ described a non-Hermitian topological insulator, which can be obtained by introducing nonreciprocal hopping into a Bernevig-Hughes-Zhang (BHZ) model~\cite{Science2006BHZ,Science2014Qian_Quantum,Science2018Wu_obsevation,PRL2018ZBY_Majorana}. 
$\delta_{x/y}=(t^{-}_{x/y}-t^+_{x/y})/2$ and $t_{x/y}=(t^{-}_{x/y}+t^{+}_{x/y})/2$ with $t_{x/y}^{\pm}$ denotes the nonreciprocal hopping in $x/y$ direction, respectively. $\Delta(\mathbf k)=  \Delta_x\cos k_x+ \Delta_y\cos k_y$ is pairing term with $\Delta_x=-\Delta_y$ for $d$-wave pairing. 
The Bloch Hamiltonian satisfies PHS with $\tau_x H^{T}(k_x, k_y)\tau_x = -H(-k_x, -k_y)$ and TRS with $s_y H^*(k_x, k_y) s_y = H(-k_x, -k_y)$.    
                                                             	
To elucidate the unique properties of the non-Hermitian superconductor, we show the energy spectrum for a $30 \times 30$ square lattice in Fig.~\ref{fig_bulk}(a). 
The black and gray dots represent the energies under open boundary condition (OBC) and periodic boundary condition (PBC), respectively. The energy spectrum under OBC is quite different from that under PBC, with a notable observation of energy collapse.
The inset reveals the eigenenergies of MCMs within the energy gap.  In the Hermitian case, these MCMs are eightfold degeneracy. But the eightfold degenerate MCMs are broken into two parts in the non-Hermitian case: MCMs with pure imaginary eigenenergies and MZMs with zero eigenenergies.
The summed probability distribution of bulk states is shown in Fig.~\ref{fig_bulk}(b), where the states are localized at the two opposite corners of the open system. Specifically, the corresponding eigenstates of the particles and the holes are localized at the opposite corners of the lattice which is known as the $Z_2$ skin effect and protected by the PHS. 

\begin{figure}
\centering     
\includegraphics[width=0.45\textwidth]{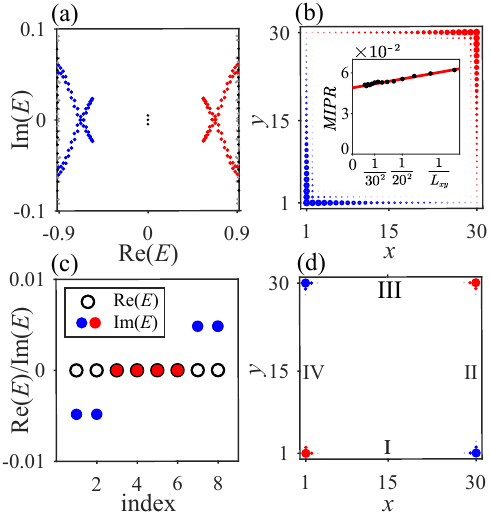}
\caption{(a) The energy spectrum around $Z_2$ skin-topological modes, marked by red and blue dots.
(b) The summed probability distributions of the $Z_2$ skin-topological modes, which are localized at opposite corners of the lattice. The inset is the size scaling of $MIPR$ for the $Z_2$ skin-topological modes.
(c) The eigenenergies of Majorana modes, with solid circles denoting the real parts and hollow circles representing the imaginary parts. 
(d) The corresponding probability distributions of MCMs with pure imaginary energies (marked in blue) and MZMs (marked in red).
Four edges are labelled \Rmnum{1}, \Rmnum{2}, \Rmnum{3} and \Rmnum{4}, which will be used in the edge theory.
Parameters are set as same as in Fig.~\ref{fig_bulk}.}
\label{fig_edge}
\end{figure}

The eigenenergies of the particles and holes of topological edge modes are shown as red and blue dots in Fig.~\ref{fig_edge}(a).
The corresponding summed probability distributions of the topological edge modes are represented by the same colors in Fig.~\ref{fig_edge}(b).
Different from the Hermitian case, the topological edge modes of corresponding particles and holes are localized at opposite corners of the lattice, showing that $Z_2$  skin-topological effect depends on the particle and hole degrees of freedom and is protected by the PHS. The $Z_2$ skin-topological modes are induced by the interplay between the topological edge modes and the $Z_2$ skin effect.
To analyze their distribution behavior, we introduce the mean inverse participation ratio ($MIPR$) of topological edge modes, which was defined by $MIPR=\frac{1}{N_L}\sum_{n=1}^{N_{L}}(\sum_{m}\abs{\psi_n^m}^4)/(\sum_{m}\abs{\psi_n^m}^2)^2$, where $N_L$ represents the number of the topological edge modes and $\psi_{n}^{m}$ denotes the density distribution of $n$th topological edge mode at site $m$. 
The size scaling of the $MIPR$ for the non-Hermitian system is shown in the inset of Fig.~\ref{fig_edge}(b). 
The $MIPR$ of the topological edge modes is inversely proportional to the size of the lattice ($L_{xy}=L^2$) and approaches a nonzero value in the thermodynamic limit, which indicates these modes will accumulate at corners of the lattice and is essentially different from the Hermitian case ($\lim_{L\rightarrow\infty} MIPR = 0$).
The eigenenergies of the MCMs are depicted in Fig.~\ref{fig_edge}(c), revealing four MCMs (marked in blue) with pure imaginary energies and four MZMs (marked in red).
The corresponding probability distributions are shown in Fig.~\ref{fig_edge}(d), revealing that the eight MCMs are sharply localized at the four corners of the square lattice. Consequently, the 2D non-Hermitian superconductor is a NHSOTSC.
Notably, the four MZMs are localized at the same two corners as the bulk states, i.e., the left-down and right-up corners, which were marked in red in Fig.~\ref{fig_edge}(d). The MCMs with pure imaginary eigenenergies are localized at the other two corners, which are marked in blue. 

The Bloch Hamiltonian characterizes the properties of the systems under PBC. However, the non-Hermitian system under OBC exhibits unique phenomena that do not have Hermitian counterparts, which indicates the breakdown of the Bloch band theory and the necessitating resorting to the non-Bloch band theory for our system.
For a $D$-dimensional Bloch Hamiltonian $\mathcal{H}(\mathbf k)$, the corresponding non-Bloch Hamiltonian $\mathcal{H}(\bm\beta)$ can be obtained by extending the real-valued wave numbers $\mathbf k$ to the complex-valued non-Bloch wave numbers $\mathbf{\tilde k}=\mathbf k -i \bm\gamma$ with $\bm\beta=e^{i\mathbf{\tilde k}}$ with $\mathbf k=(k_1,\cdots,k_D)$ and $k_j/\gamma_j\  (j=1,\cdots, D)\in\mathbb{R}$. 
The $n$th right and left non-Bloch eigenstates $\ket*{\psi_n^{R/L}}$ are defined by  $\mathcal{H}\ket*{\psi_n^R}=E_n\ket*{\psi_n^R}$ and $\mathcal{H}^{\dagger}\ket*{\psi_n^L}=E_{n}^{*}\ket*{\psi_n^L}$, respectively, and obey the
biorthogonal normalization $\bra*{\psi^L_m}\ket*{\psi_n^R}=\delta_{m,n}$. The non-Bloch Hamiltonian satisfied non-Bloch PHS with $C \mathcal{H}^T(\bm\beta)C^{-1} = -\mathcal{H}(\bm\beta^{-1})$, $CC^{*}=\pm 1$. Here we take the notation $\bm\beta^{-1}=(\beta_1^{-1},\cdots,\beta_D^{-1})$.
The values of $\bm\beta$, which formed the generalized Brillouin zone (GBZ), can be determined by solving the characteristic equation:
\begin{equation}
f(\bm\beta, E) = \det[\mathcal{H}(\bm\beta)- E] = 0.
\label{eq_CE}
\end{equation} 
Note that $f=f(\bm\beta, E) $ is a Laurent polynomial of $\beta_j$ for a given energy $E$.  
Here, we reveal the properties of the non-Bloch bands by the Amoeba theory~\cite{wang2022amoeba}, which provides a general framework for studying the non-Bloch bands beyond one-dimensional non-Hermitian systems. The Amoeba~\cite{wang2022amoeba,ComputingAmoebas} of $f$ is defined as the logarithm modulus of the zero locus of $f$:
\begin{equation}
\mathcal A_f=\{(\log{\abs{\beta_1}},\cdots,\log{\abs{\beta_D}}):f(\bm\beta, E)=0\}.
\label{eq_amoeba}
\end{equation} 
The Ronkin function~\cite{wang2022amoeba,Amoeba2} of $f$ is defined as: 
\begin{equation}
R_f(\bm\gamma, E)=\frac{1}{(2\pi)^D}\int_{T^D}\dd k^D \log\abs{f( \bm\beta, E)},
\label{eq_ronkin}
\end{equation}
where $T^D=[0,2\pi]^D$ is a D-dimensional torus and $\dd k^D=\dd k_1 \dd k_2 \cdots \dd k_D$. The gradient of the Ronkin function denoted by $\nu_{j}=\frac{\partial R_f}{\partial \gamma_{j}}$ is related to the winding number:
\begin{equation}
w_{j}=\frac{1}{2\pi i}\int_0^{2\pi}\dd k_{j}\frac{\partial_{k_{j}} f(\bm\beta, E)}{ f(\bm\beta, E)},
\label{eq_windn}
\end{equation} 
with: 
\begin{equation}
\nu_{j}(\bm\gamma,E)=\int_{T^{D-1}}\frac{\dd k_{1}\cdots\dd k_{j-1}\dd k_{j+1}\cdots\dd k_{D}}{(2\pi)^{D-1}} w_{j}
\label{eq_grad}.
\end{equation} 
In the thermodynamic limit, the minimum of the Ronkin function $(\bm\nu=0)$ yields a singular value $R_{f}(\bm\gamma^\text{min}, E)$  when the energy $E$ lies within the OBC energy spectra. 
Therefore,  $\gamma_{j}^\text{min}=\log |\beta_{j}|$ determines both the decay factor of eigenstates and the shape of GBZ. 
Due to the PHS, when  $\mathcal A_f$ is a Amoeba  for $f(\bm\beta, E) $ with an eigenvalue $E$, we have
\begin{align}
f(\bm\beta, E) &= \det[\mathcal{H}(\bm\beta)-E]\nonumber
\\&=\det[-C \mathcal{H}^T(\bm\beta^{-1})C^{-1}-E] \nonumber
\\&=\det[\mathcal{H}(\bm\beta^{-1})+E]\nonumber
\\&=f(\bm\beta^{-1}, -E)=0.
\label{eq_CE2}
\end{align}
Thus, the Ronkin function and its gradient of particle and hole satisfy  $R_f(\bm\gamma ,E)=R_{f}(-\bm\gamma,-E)$ and $ \nu_{j}(\bm\gamma ,E)=\nu_{j}(-\bm\gamma,-E)$. 
Therefore, we have
\begin{align}
\bm\gamma^\text{min}(E)=-\bm\gamma^\text{min}(-E),
\label{eq_gamma}
\end{align}
which indicates that $\gamma_j^\text{min}$ and $-\gamma_j^\text{min}$ are always paired for a paired particle ($E$) and hole ($-E$). If the eigenstates localized at one boundary for $\gamma_j^{\text{min}}<0$, they will localized at the opposide boundary for  $\gamma_j^{\text{min}}>0$. Consequently, the corresponding eigenstates of particles and holes are localized at the opposite boundaries of the lattice, recognized as $Z_2$ skin effect and protected by the PHS.
For example, in our non-Hermitian $d$-wave pairing superconductor, the eigenstates localize at the left/down boundary of the lattice when $\gamma_x^\text{min}/\gamma_y^\text{min}<0$, and at the right/up boundary when $\gamma_x^\text{min}/\gamma_y^\text{min}>0$. 
The corresponding eigenstates of particles and holes are localized at the opposite corners of the lattice, as shown in Fig.~\ref{fig_bulk}(b).

Note that the non-Hermitian skin effect exists in two and higher-dimensional systems if the PBC energy spectrum covers a finite area~\cite{zhang2022universal}. 
The non-Hermitian skin effect can be classified by the current function based on the PBC energy spectrum. 
For a band $m$, the current function takes the form
\begin{align}
	J^m_{\alpha}[n]=\oint_\mathrm{BZ}\dd k^{D} n(E_m,E_m^*)\nabla_{\alpha} E_m(k),
\end{align}
in which $\nabla_\alpha$ indicates the directional derivative along certain $\alpha$ in $D$-dimensional Brillouin zone. 
$n(E, E^*)$ denotes the distribution function in a steady state and only depends on the energy. For example, the Fermi distribution satisfies $n(E,E^*)=1/[1+e^{\Re(E)/k_BT}]$. Due to the PHS, the energies of particles ($E_p$) and holes ($E_h$) occur in pairs with the relation of $E_p(k)=-E_h(-k)$ under PBC. 
For a superconductor, the distribution function satisfied $n(E_p,E_p^*)=1-n(E_h,E_h^*)$. 
Then, the current functions for the bands of particles ($J^p_{\alpha}[n]$) and holes ($J^h_{\alpha}[n]$) are always opposite due to
\begin{align}
	J^p_{\alpha}[n]&=\oint_\mathrm{BZ}\dd k^{D} n(E_p,E_p^*)\nabla_{\alpha} E_p(k)
	\nonumber \\
	&=\oint_\mathrm{BZ}\dd k^{D} [1-n(E_h,E_h^*)]\nabla_{-\alpha} E_p(-k)
	\nonumber \\
	&=-\oint_\mathrm{BZ}\dd k^{D} n(E_h,E_h^*)\nabla_{\alpha} E_h(k)
	\nonumber \\
	&=-J^h_{\alpha}[n],
\end{align}
which also indicates that the non-Hermitian skin effect of a superconductor is $Z_2$ skin effect and belongs to the framework of reciprocal skin effect~\cite{zhang2022universal}. 

In addition, the non-Hermitian skin effect can be characterized by the point-gap topology of the non-Hermitian systems~\cite{KawabataPRX2019,lin2023topological,ding2022non,prb2020higherorder}.
In one-dimensional systems, The non-trivial winding number of the Bloch Hamiltonian indicates the emergence of the non-Hermitian skin effect~\cite{OkumaPhysRevLett2020,zhangPRL2020}. For our 2D non-Hermitian superconductor, the $Z_2$ skin effect and the $Z_2$ skin-topological effect can also be uncovered by the point-gap topology in a cylinder geometry.
The PHS induces the spectrum winding numbers of particles and holes to be opposite in both the bulk states and topological edge modes  in the cylinder geometry, revealing the $Z_2$ skin effect and $Z_2$ skin-topological effect (see appendix for more details).
\section{Topological Phase transition at Bloch points}

In the previous section, we have shown that the non-Hermitian system can stabilize the topologically nontrivial NHSOTSCs with robust in-gap MCMs. The topological phase transition between the topologically nontrivial and trivial phases occurs when the gap closes. 
In one-dimensional non-Hermitian superconductor systems, the energy gap only closes at the Bloch points~\cite{JPCM2023Yang_Universal}. 
Here, we extend it into two and higher-dimensional non-Hermitian superconductors by the non-Bloch band theory with the Amoeba formulation to reveal the topological phase transition.
From Eq. \ref{eq_gamma}, when the gap of bulk's energies close with $E=0$, we can deduce that
\begin{align}
	\gamma_{j}^{\mathrm{min}}(0)=-\gamma_{j}^{\mathrm{min}}(0)=0,\quad  j=1,\cdots, D.
\end{align}
Therefore, the energy gap closes at the Bloch points which are the intersections of the GBZs of particles and holes and the Brillouin zone with
\begin{align}
	|\beta^{BP}_{j}| = e^{\gamma_{j}^\text{min}} = 1,\quad j=1,\cdots,D.
	\label{eq:BP}
\end{align}
Thus, the topological phase transitions occur when the gap closes at the Bloch points. Therefore, the phase transition points determined by the Bloch Hamiltonian and the non-Bloch Hamiltonian coincide with each other for non-Hermitian superconductors with the PHS. 
This provides a useful approach for exploring topological phase transitions in higher-dimensional superconductors. 

\begin{figure}
	\includegraphics[width=0.45\textwidth]{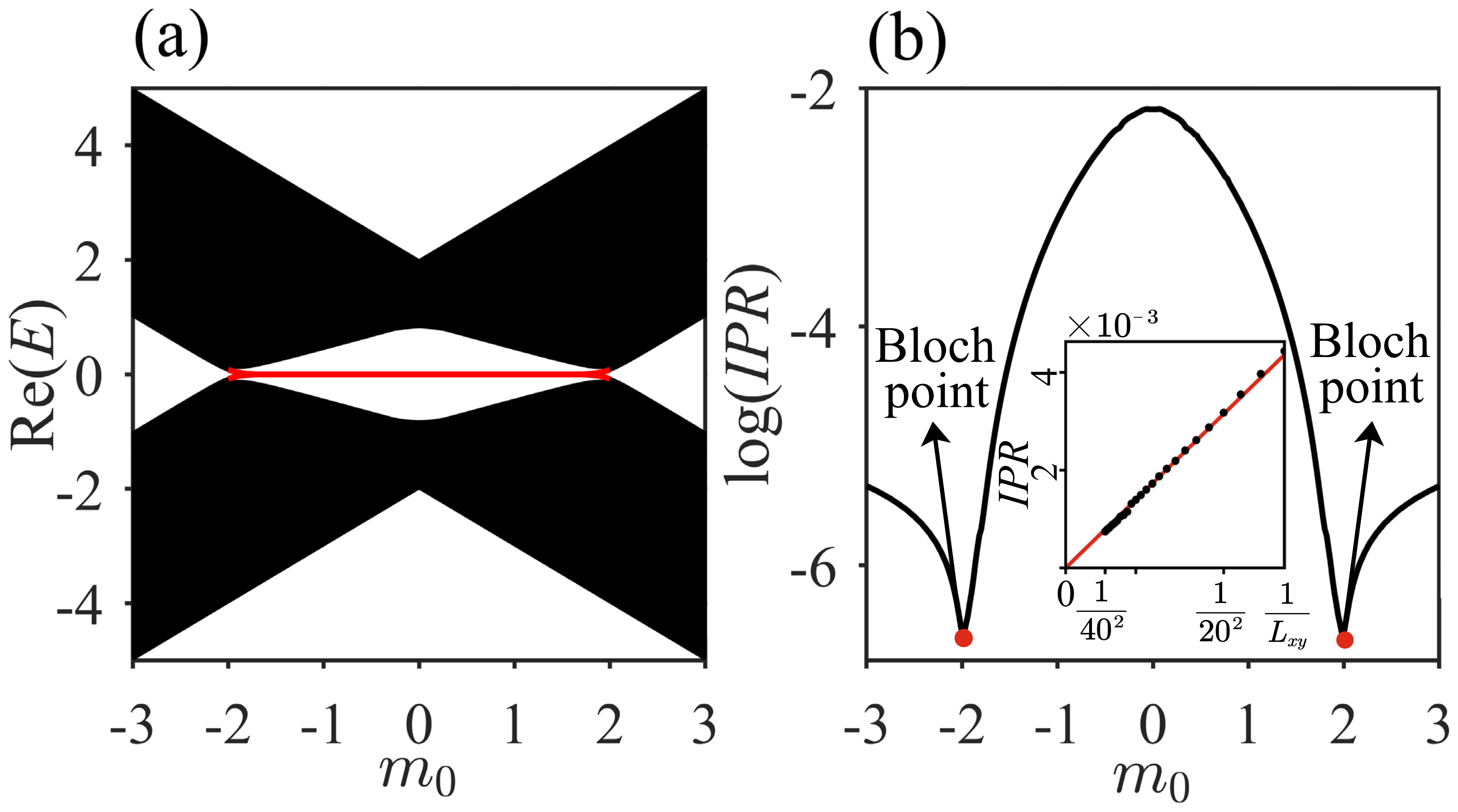}
	\caption{(a) The real parts of the energy spectrum vary with $m_0$, resulting from a $30\times 30$ square lattice with open boundaries along both $x$ and $y$ directions. The energy gap closes at $m_0=\pm (t_x+t_y)$. (b) Numerical results of $\log(IPR)$ with varied $m_0$, extracted from the states possessing the minimal energy modulus. The value of $\log(IPR)$ is minimal at the Bloch point. The inset details the $IPR$ of the state of Bloch point, presented as a function of lattice size $L_{xy}=L\times L$.
	Other parameters are $t_x=t_y=1$, $A_x=A_y=1$, $\mu=0$, $\delta_x = \delta_y = 0.7$ and $\Delta_x = -\Delta_y = 0.5$.}\label{fig_BPs}
\end{figure}

For the sake of simplicity, we focus on the $\mu=0$ and $t_{x/y}>0$ case for our systems in the subsequent analysis, where simple analytic results for topological edge modes can be obtained.
Fig.~\ref{fig_BPs}(a) shows the real part of the OBC energy spectrum as a function of $m_0$. The topological phase transition occurs when the gap closes at critical values $m_0 = \pm(t_x + t_y)$ with $E = 0$. The phase is topologically nontrivial with robust MCMs when $|m_0| < t_x + t_y$. The gap closes at the Bloch points with $|\beta_{x/y}|=1$.
In our model, the non-Bloch Hamiltonian has an additional TRS $s_y H^*(\beta_x,\beta_y)s_y = H(\beta_x^{-1},\beta_y^{-1})$, which requires an additional condition of gap closing points with $\beta_{x/y} =\beta_{x/y}^{*}$. 
Then, the gap closes at critical values $m_0 = t_x+t_y$ with $(\beta^{BP}_{x}$, $\beta^{BP}_{y}) = (1,1)$ and $m_0 = -(t_x+t_y)$ with $(\beta^{BP}_{x}$, $\beta^{BP}_{y}) = (-1,-1)$, respectively. The eigenstates of the Bloch points are extended states without any localization, which can be used to determine the phase transitions in experiments. To confirm the characteristic of the Bloch points, we calculate the logarithm of the inverse participation ratio ($IPR$), defined by $\log (IPR) = \log (\sum_{i=1}^N|\psi_i|^4/| \sum_{i=1}^N|\psi_i|^2|^2)$,
which represents the intensity of state localization. 
In the thermodynamic limit, the value of $\log(IPR)$ approaches zero when a state becomes increasingly localized, while it tends towards negative infinity for an extended state.

Fig.~\ref{fig_BPs}(b) represents the value $\log (IPR)$ of the states with the minimal modulus of energy as a function of the parameter $m_0$. The $\log(IPR)$ reaches minimal values when the energy gap closes at the Bloch points. The insert shows the size scaling of the $IPR$ of the Bloch points. The value of $IPR$ tends towards zero as the lattice size increases, which numerically confirms that the eigenstates of the Bloch points are extended waves without any localization.

\section{Edge Theory}
To gain an intuitive understanding of the MCMs, we adopt the edge theory~\cite{PRL2018ZBY_Majorana,PRL2019TL_second} in the non-Hermitian system. 
From Eq. \ref{eq:BP}, we know that the energy gap closes at the Bloch points with real-valued wave numbers. We consider the low-energy case by expanding the Hamiltonian up to the second-order of momentum around the Bloch point with ($\beta^{BP}_{x}$, $\beta^{BP}_{y}) = (1, 1)$:
\begin{align}
H_0(\mathbf k)&=(m+\frac{t_x}{2}k_x^2+\frac{t_y}{2}k_y^2)\tau_z\sigma_z  \nonumber- \frac{1}{2}(\Delta_x k_x^2 + \Delta_y k_y^2)\tau_ys_y 
\\&  + A_x k_x s_z\sigma_x+ A_y k_y\tau_z\sigma_y - i\delta (k_x+k_y)\sigma_z,		
\end{align}
with $m=m_0-(t_x+t_y) <0$. To streamline our analysis, we designate the four edges of the square lattice as \Rmnum{1}, \Rmnum{2}, \Rmnum{3}, and \Rmnum{4} as illustrated in Fig.~\ref{fig_edge}(d). 
For edge \Rmnum{1}, we can substitute $k_x$ with $-i\partial_x$ and disregard the insignificant $k_y^2$ term. Considering scenarios where the pairing term is relatively small, $H_0$ can be decomposed into two parts, $H_1$ and $H_2$, where $H_2$ can be regarded as a perturbation:
\begin{align}
H_1(-i\partial_x,k_y) &=(m-t_x \partial_x^2/2)\tau_z\sigma_z - iA_x s_z\partial_x ,\nonumber
\\ H_2(-i\partial_x,k_y) &= A_y k_y\tau_z\sigma_y + \frac{\Delta_x}{2}\tau_y s_y\partial_x^2 
- \delta(\partial_{x}+ik_y)\sigma_z. 
\end{align}

By solving the eigenvalue equation $H_1\ket{\psi_1}=E_1\ket{\psi_1}$  under the boundary condition $\psi_1(0) = \psi_1(\infty) = 0$, we obtain the four Jackiw-Rebbi solutions:
\begin{equation}
	\psi^\text{\Rmnum{1}}_\alpha(x)=N_x\sin(\kappa_1 x)e^{-\kappa_2 x}e^{ik_y y}\chi_\alpha^\text{\Rmnum{1}},
\end{equation}
with $\kappa_1=\sqrt{-2m/t_x-A_x^2/t_x^2}$ and $\kappa_2=A_x/t_x$. The normalization coefficient is given by $N_x=2\sqrt{\kappa_2(\kappa_1^2+\kappa_2^2)/\kappa_1^2}$. 
The normalization of the wave function requires $\kappa_1^2+\kappa_2^2>0$, which in turn implies that $m_{0}<t_x+t_y$. 
Consequently, the existence of edge modes directly corresponds to whether the phase is topologically nontrivial, which aligns with the generalized bulk-boundary correspondence.
The eigenvector $\chi^{\text{\Rmnum{1}}}_{\alpha}$ satisfy 
$\tau_z s_z\sigma_y\chi^\Rmnum{1}_\alpha=-\chi_\alpha^\Rmnum{1}$ 
, and can be chosen as:
\begin{eqnarray}
	\chi^{\text{\Rmnum{1}}}_1=\ket{\tau_z=+1}\otimes \ket{s_z=+1} \otimes \ket{\sigma_y=-1}, \nonumber
	\\ \chi^\text{\Rmnum{1}}_2=\ket{\tau_z=+1}\otimes \ket{s_z=-1}\otimes \ket{\sigma_y=+1},\nonumber
	\\ \chi^\text{\Rmnum{1}}_3=\ket{\tau_z=-1} \otimes \ket{s_z=+1} \otimes \ket{\sigma_y=+1},\nonumber
	\\ \chi^\text{\Rmnum{1}}_4=\ket{\tau_z=-1} \otimes \ket{s_z=-1} \otimes \ket{\sigma_y=-1}.
\end{eqnarray}

The matrix element of $H_{\text{edge}}^\text{\Rmnum{1}}$ in this basis are: 
$H_{\text{edge}}^{\text{\Rmnum{1}},\alpha\beta}(k_y) = \int_0^{\infty}\dd x \psi^{\text{\Rmnum{1}}*}_\alpha H_2\psi^\text{\Rmnum{1}}_\beta$. 
Therefore, we get the effective Hamiltonian of edge \Rmnum{1}($H_{\text{edge}}^\text{\Rmnum{1}}$).
For the same as other edges, the effective $4\times 4$ edge Hamiltonian of four edges can be written as: 
\begin{eqnarray}
H_{\text{edge}}^\text{\Rmnum{1}}(k_y)=&-A_yk_ys_z+(\Delta_xm/t_x)\tau_y s_y, \nonumber
\\H_{\text{edge}}^\text{\Rmnum{2}}(k_x)=&A_xk_xs_z+(\Delta_ym/t_y)\tau_y s_y, \nonumber
\\H_{\text{edge}}^\text{\Rmnum{3}}(k_y)=&A_yk_ys_z+(\Delta_xm/t_x)\tau_y s_y, \nonumber
\\H_{\text{edge}}^\text{\Rmnum{4}}(k_x)=&-A_xk_xs_z+(\Delta_ym/t_y)\tau_y s_y.
\end{eqnarray}
The effective Hamiltonian of the edges can be represented by considering the edge coordinate $l$ in an anticlockwise direction. The edge Hamiltonian can be rewritten as $H_\text{edge}=-iA(l)s_z\partial_{l}+M(l)\tau_y s_y$. The coefficients, $A(l)$ and $M(l)$, are step functions with particular values: $A(l)=A_y$, $A_x$, $A_y$, $A_x$, 
and $M(l)=\Delta_x m/t_x$, $\Delta_y m/t_y$, $\Delta_x m/t_x$ and $\Delta_y m/t_x$, respectively.

Due to $d$-wave pairing ($\Delta_x=-\Delta_y$), the mass term of the effective Hamiltonian undergoes sign changes at the corners. 
As a result, one Majorana Kramers pair manifests at each corner, akin to the Jackiw-Rebbi zero modes protected by the PHS~\cite{prd1976jackiw}. 
It is worth noting that the NHSOTSC is not limited to $d$-wave pairing.
The fully gapped $s_\pm$-wave paired superconductor with the pairing term $\Delta(\mathbf k)=\Delta_0-\Delta_s(\cos k_x+ \cos k_y)$ ($0 < \Delta_0 < \Delta_s$) can be topologically nontrivial NHSOTSC when the mass term of the edge Hamiltonian alternates in sign at each corner, which guarantees the existence of the topological robust MCMs.

\section{$p+ip$-wave pairing SOTSC}

\begin{figure}
	\includegraphics[width=0.45\textwidth]{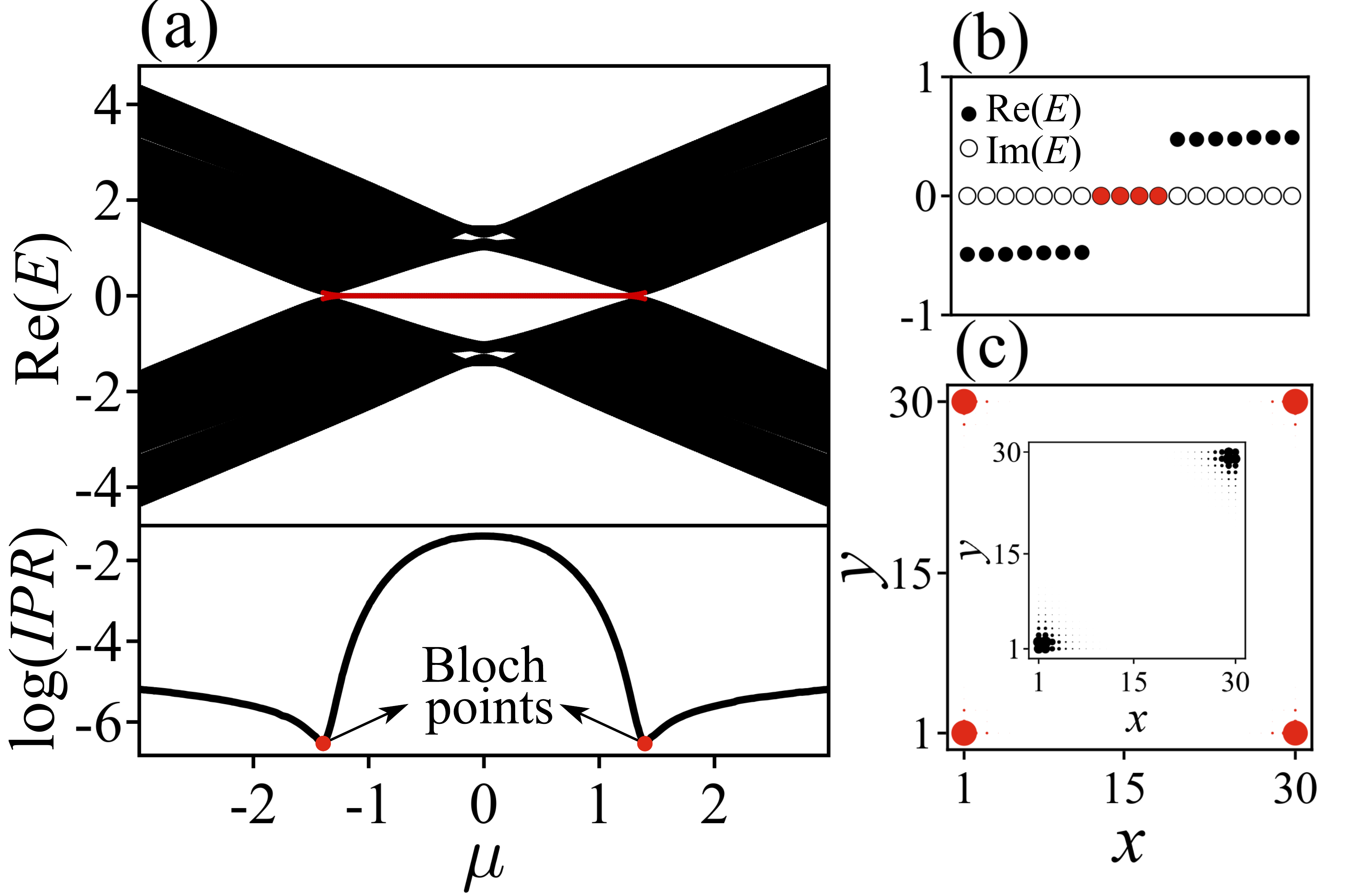}
	\caption{ (a) The top figure illustrates the variation in energy spectra of a $30 \times 30$ square lattice under OBC with $\mu$. The topological MCMs are denoted by the red dots. The bottom figure showcases the $\log(IPR)$ of the states that possess the smallest energy modulus. (b) The eigenenergies around the gap with $\mu=0.8$. (c) The summed probability distribution of the MCMs and the bulk states(inset). Parameters are set as $t=1$, $\delta=0.2$ and $\Delta=1$.
	}
	\label{fig_pwave}
\end{figure}

Having delved into the 2D NHSOTSC model with $d$-wave pairing, it's noteworthy that our methodology for achieving non-Hermitian higher-order superconductors isn't confined to $d$-wave pairing. Moreover, the characteristics of non-Hermitian superconductors rely solely on PHS, irrespective of the type of pairing or the number of bands. We now put forth a prototypical 2D NHSOTSC model that incorporates $p+ip$-wave pairing~\cite{PRR2020Tiwari_Chiral,PRB2018wangWeek}, showcasing another substantial pairing mechanism. The Bloch Hamiltonian is written as follows:
\begin{eqnarray}
H(\mathbf k)=
t\cos{k_x}\tau_z\sigma_x+t\cos{k_y}\tau_z\sigma_z
+\Delta\sin{k_x}\tau_x \nonumber
\\+\Delta\sin{k_y}\tau_y-\mu\tau_z-i\delta\sin{k_x}\sigma_x-i\delta\sin{k_y}\sigma_z.
\end{eqnarray}                 
The Hamiltonian preserves the PHS with $\tau_x H(\mathbf{k})^{T}\tau_x=-H(-\mathbf{k})$.
The non-Bloch Hamiltonian exhibits an additional symmetry $H^\dagger(\beta_x,\beta_y)=\sigma_z H(\beta_x^*,\beta_y^*)\sigma_z$, necessitating a further condition on the critical values $\beta^c_{x/y}$, given as $\beta_{x/y}^c=\beta_{x/y}^{c*}$. Consequently, the conditions for energy gap closure become $\beta_{x/y}^c=\pm 1$, that is, $ (\tilde{k}_x, \tilde{k}_y)=(0,0), (0, \pi), (\pi, 0), (\pi, \pi)$. Thus, the gap-closing points occur when $\mu=\pm\sqrt{t_x^2+t_y^2}$. 
The upper panel of Fig.~\ref{fig_pwave}(a) illustrates the real part of the energy spectrum as functions of $\mu$. 
The topological region is distinctly delineated in red dots for the topological robust MCMs. 
The lower panel of Fig.~\ref{fig_pwave}(a) gives the $\log (IPR)$ as a function of $\mu$ for the states possessing minimal energy modulus.
The $\log(IPR)$ reaches its minimum at the gap-closing points and approaches negative infinity in the thermodynamic limit, a hallmark of the extended nature of these states and a unique contribution of Bloch points. 
Fig.~\ref{fig_pwave}(b) shows the eigenenergies around zero for $\mu=0.8$, revealing the presence of four MZMs within the gap.
The probability distribution of the MZMs, as depicted in Fig.~\ref{fig_pwave}(c), reveals a sharp localization of MCMs at the four corners of the lattice — a characteristic feature of the second-order topological phase in 2D systems. As shown in the inset of Fig. \ref{fig_pwave}(c), bulk states are localized at opposite corners of the lattice, signifying the $Z_2$ skin effect.

\section{Conclusion}
In summary, we have uncovered the topological properties of NHSOTSC in nonreciprocal systems with $d$-wave and $p$-wave pairing, respectively. Due to the PHS, the bulk eigenstates exhibit $Z_2$ skin effect. The topological edge modes are localized at the opposite corners, indicating the $Z_2$ skin-topological effect depends on the particle and hole degrees of freedom. Contrary to the eightfold degenerate MZMs in the Hermitian case, the MCMs in the non-Hermitian system consist of two parts: MCMs with pure imaginary energies and MZMs. The four MZMs are localized at the same corners as the bulk states, while the four MCMs with pure imaginary energies are localized at the other two corners. 
We have elucidated the MCMs and topological phase transitions in the framework of non-Bloch band theory with the Amoeba formulation. The topological phase transition occurs only when the gap closes at the Bloch points, which provides a convenient and effective method to determine the topological phase transitions. Based on edge theory, we have intuitively uncovered the emergence of topologically nontrivial MCMs and established the generalized bulk-boundary correspondence for the NHSOTSCs. Our study provides a framework to explore the richer non-Hermitian phenomena in higher-order topological superconductors. 

\section{Acknowledgments}
We thank Prof. Z. Wang and Z. Yan for fruitful discussions. This work was supported by Natural Science Foundation of Jiangsu Province (Grant No. BK20231320) and National Natural Science Foundation of China (Grant No. 12174157).

\appendix

\section*{Appendix: Point-gap topology of $Z_2$ skin effect and $Z_2$ skin-topological effect}

We have investigated the $Z_2$ skin effect of the superconductors by non-Bloch band theory based on GBZ and the current function based on the PBC energy spectrum. Here, we uncover the $Z_2$ skin effect and $Z_2$ skin-topological effect by the point-gap topology of the system with the cylinder geometry. 

\begin{figure}[htbp]
	\includegraphics[width=8.4cm]{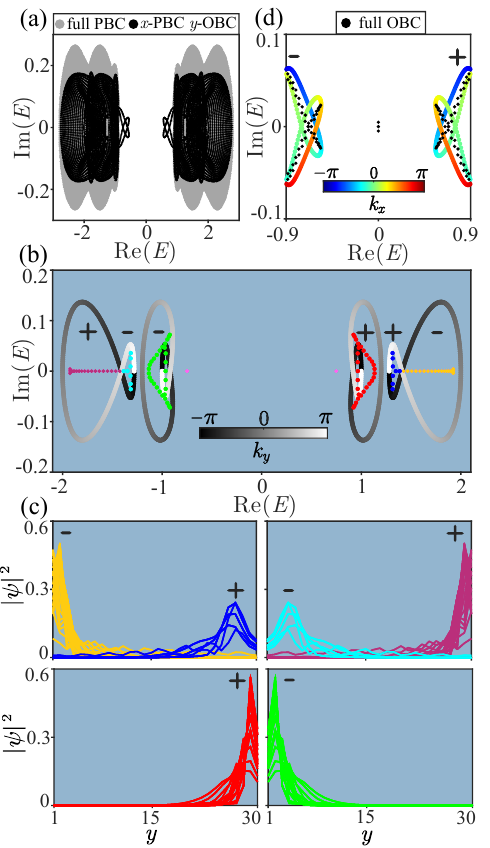}
	\caption{The energy spectra and probability distributions of the non-Hermitian superconductor in the main text [Eq. (\ref{eq_1})]. 
			(a) Energy spectra of the system under full PBC (gray dots) and only OBC in $y$ direction (black dots).
			(b) Energy spectrum for a fixed $k_x=0$.
			The grayscale points represent the spectrum under $y$-PBC and colored points denote the energies under $y$-OBC. 
			(c) The probability distribution of energies that correspond to the energies denoted by the points with the same colors in (b).
			(d) Energy spectrum around the $Z_2$ skin-topological modes under full OBC (black dots) and only OBC in $y$ direction (colorful dots).
			The lattice length in $y$-directional is $L_y=30$. Other parameters are the same as Fig. $\ref{fig_bulk}$ in the main text.}
	\label{fig_A1}
\end{figure}

The Hamiltonian in Eq. (\ref{eq_1}) under PBC in $x$ direction ($x$-PBC) and OBC in $y$ direction ($y$-OBC) share the same spectrum as the one under $y$-PBC and $x$-OBC. 
Thus, we investigate the system under the cylinder geometry with $x$-PBC and $y$-OBC to reveal the non-Hermitian skin effect in $y$ direction without loss of the generality.
The complex energy spectra of the Hamiltonian in Eq. (\ref{eq_1}) under PBC and cylinder geometry are shown in Fig. \ref{fig_A1} (a) with gray and black dots, respectively. 
The PBC energy spectrum covers a finite area while the energy spectrum of cylinder geometry collapses, which indicates the existence of non-Hermitian skin effect.

To characterize the non-Hermitian skin effect by the point-gap topology, we show the energy spectrum of the full PBC and $y$-OBC in Fig. \ref{fig_A1}(b) for a fixed $k_x=0$ with grayscale and colorful points, respectively. 
The point-gap topology can be characterized by the winding numbers $W(k_x,E)$ for a reference energy $E$, which was defined by 
\begin{align}
	W(k_x,E)=\oint_{0}^{2\pi}\frac{\dd k_y}{2\pi i}\frac{\partial}{\partial k_y}\log \det [H(k_x,k_y)-E].
\end{align}
The positive/negative winding number corresponds to the anti-clocked/clocked direction of the energy variation with the momentum.
The sign of the winding number for each band is marked by $+/-$ in Fig. \ref{fig_A1}(b). The winding number of the band for particles is always opposite to the one for the corresponding holes with $W(k_x,E)=-W(-k_x,-E)$:
\begin{align}
&W(k_x,E)
\nonumber \\
=&\oint_{0}^{2\pi}\frac{\dd k_y}{2\pi i}\frac{\partial}{\partial k_y}\log \det [H(k_x,k_y)-E]
\nonumber \\
=&\oint_{0}^{2\pi}\frac{\dd k_y}{2\pi i}\frac{\partial}{\partial k_y}\log \det [-\tau_x H(-k_x,-k_y)\tau_x-E]
\nonumber \\
=&-\oint_{0}^{2\pi}\frac{\dd k_y}{2\pi i}\frac{\partial}{\partial k_y}\log \det [ H(-k_x,-k_y)+E]
\nonumber \\
=&-W(-k_x,-E).
\end{align}

The corresponding eigenstates of energies in Fig. \ref{fig_A1}(b) are depicted in Fig. \ref{fig_A1}(c) with the same colors. The non-zero winding numbers of PBC energies for a fixed $k_x$ indeed indicates the emergence of the non-Hermitian skin effect in $y$ dimension. Furthermore, the positive winding numbers correspond to the accumulation of eigenstates at the right boundary of $y$ direction, while the negative winding numbers represent the accumulation of the eigenstates at the opposite boundary. Thus, the eigenstates of the particles and holes accumulate at opposite boundaries with opposite winding numbers, resulting in the $Z_2$ skin effect which is protected by the PHS. The non-Hermitian skin effect in $x$ dimension can also be characterized by the winding numbers $W(k_y, E)$ the same as the above.

The $Z_2$ skin-topological effect, which is the interplay between the topological edge modes and $Z_2$ skin effect, can also be revealed by the point-gap topology qualitatively.
The energies of the topological edge modes under the cylinder geometry (colorful dots) encircle the energies of the skin-topological modes under full OBC (black dots) as depicted by Fig. \ref{fig_A1}(d), revealing that the non-Hermitian skin effect acts on the topological edge modes. 
Due to the PHS, the winding numbers of particles and holes are opposite under the cylinder geometry, which indicates the skin-topological modes of particles and the holes are localized at opposite boundaries as shown in Fig.~\ref{fig_edge}(b). 
Consequently, the skin-topological effect in a non-Hermitian superconductor system corresponds to the $Z_2$ skin-topological effect. In addition, the gap in the energies of topological edge modes under cylinder geometry provides a platform for the emergence of second-order topological corner modes.

\bibliographystyle{apsrev4-2}
\bibliography{reference} 
	
\end{document}